\documentclass[epj]{svjour}
\usepackage{graphics}
\usepackage{hyperref}
\newcommand{\ys}{{\rm Y_S}}
\newcommand{\rcz}{{\rm R_{CZ}}}
\newcommand{\msun}{{\rm M_\odot}}
\newcommand{\onemsun}{1\,$\msun$}
\newcommand{\rsun}{{\rm R_\odot}}
\newcommand{\lsun}{{\rm L_\odot}}
\newcommand{\zxsun}{{\rm (Z/X)_\odot}}
\newcommand{\xini}{{\rm X_{ini}}}
\newcommand{\yini}{{\rm Y_{ini}}}
\newcommand{\zini}{{\rm Z_{ini}}}
\newcommand{\amlt}{\alpha_{\rm MLT}}
\newcommand{\tsun}{\tau_\odot}

\newcommand{\mnras}{Monthly Notices of the Royal Astronomical Society}
\newcommand{\apj}{The Astrophysical Journal}
\newcommand{\apjs}{The Astrophysical Journal Supplement Series}
\newcommand{\apjl}{The Astrophysical Journal Letters}
\newcommand{\aap}{Astronomy \& Astrophysics}
\newcommand{\araa}{Annual Review of Astronomy \& Astrophysics}
\newcommand{\solphys}{Solar Physics}
\newcommand{\nat}{Nature}
\newcommand{\prd}{Physical Review D}

\begin{document}
\title{Alive and well: a short review about standard solar models}
\author{Aldo Serenelli\inst{1}}                     
\institute{Instituto de Ciencias del Espacio (ICE/CSIC-IEEC), Campus UAB, Carrer de Can Magrans S/N, Cerdanyola del Valles, 08193, Spain}
\date{Received: date / Revised version: date}
%
\abstract{Standard solar models (SSMs) provide a reference framework across a number of research fields: solar and stellar models, solar neutrinos, particle physics the most conspicuous among them. The accuracy of the physical description of the global properties of the Sun that SSMs provide has been challenged in the last decade by a number of developments in stellar spectroscopic techniques. Over the same period of time, solar neutrino experiments, and Borexino in particular, have measured the four solar neutrino fluxes from the pp-chains that are associated with 99\% of the nuclear energy generated in the Sun. Borexino has also set the most stringent limit on CNO energy generation, only $\sim 40\%$ larger than predicted by SSMs. More recently, and for the first time, radiative opacity experiments have been performed at conditions that closely resemble those at the base of the solar convective envelope. In this article, we review these developments and discuss the current status of SSMs, including its intrinsic limitations. 
} 
\maketitle
\section{Introduction}
\label{intro}

The concept behind standard solar models (SSMs) is that of a well-defined framework within which a physical description of the Sun can be constructed and predictions be made. SSMs is, as any other model, a simplified description of nature. However, and despite the many approximations used, the SSM has been quite successful in describing many properties of the Sun. During 30 years the solar neutrino problem led to many to consider modifications to the SSMs  as a possible way to reduce the $^8$B neutrino flux to bring it into agreement with solar neutrino experiments. However, it was the excellent agreement between SSM predictions about the internal solar structure and inferences from helioseismic inversions \cite{jcd:1996,bahcall:1998} what finally convinced (almost) everyone that the solution had to be found on the particle physics sector. 

After the  final solution of the  solar neutrino problem with  the initial SNO
results \cite{SNOI}, the  role of the SSM in neutrino  physics has changed and
it  is  now   important  for  constraining  the   electron  neutrino  survival
probability,  particularly  in  the  low-energy range.  In  this  respect,  we
recommend   the    work   summarizing    the   Borexino   Phase    I   results
\cite{borexino:review} and also \cite{oscil:epj} in this Topical Issue.

The SSM also plays a fundamental role for stellar models. This is most evident when considering that convection theories used to model stars still rely on a free parameter that is calibrated by forcing solar models to reproduce the present day solar radius and temperature. Also, the calibrated initial solar composition is often used as an anchor point, together with results from Big Bang Nucleosynthesis, to link the evolution of metals and helium in the Universe. The SSM is also used as a benchmark against which we can test additional physical processes in stars. Just to mention a few examples: impact of extra mixing in Li depletion, transport of angular momentum and its impact on internal structure, core overshooting during early evolution of low mass stars.

The concordance between solar models and inferences on solar properties from helioseismology has been altered during the last decade or so, with the advent of new spectroscopic determinations of the solar surface (photospheric) composition, particularly of the abundant volatile elements C, N and O \cite{agss09}. These determinations are based on a number of qualitative advances: 3D radiation hydrodynamic  solar atmosphere models, better atomic data and NLTE-line transfer calculations.  Although not without controversy \cite{caffau:2011}, the newly determined solar abundances are lower by a large margin (30-40\%) than before \cite{gn93,gs98}. Adopting these new abundances has proven challenging for SSMs because the agreement with helioseismic data that existed previously has been lost \cite{bahcall:2005,montalban:2004,turck:2004}. Tracing an analogy with the older solar neutrino problem, the discrepancy in helioseismic results has been dubbed the solar abundance problem. This analogy also extends into that a number of modifications have been suggested to SSMs in order to resolve this problem, but all have so far failed or, at best, offered only partial limited success. 

Does the solar abundance problem set the limit to SSMs as a physical description of the Sun and, by extension, to standard stellar models? This a question that has so far no clear answer. What seems clear, however, is that the \emph{low metallicity} solar abundances are here to stay. Maybe solar CNO abundances are not as low as suggested initially by Asplund and collaborators \cite{ags05} and maybe not even as the later revised values \cite{agss09} that are currently used as  standard. But it does not seem probable that further work on solar spectroscopy will bring CNO abundances back up to values found in older works \cite{gn93,gs98}. On the other hand, the solar abundance problem has motivated further work on physical inputs to solar models, particularly improvement of nuclear reaction rates (and uncertainties; see \cite{sfii} for a complete revision of the topic), theoretical \cite{op,opas} and experimental \cite{bailey:2015} work on radiative opacities and even on the equation of state appropriate for solar conditions \cite{baturin:2013}. Recent experimental work on opacities shows that theoretical opacity calculations, the only available source of radiative opacities for solar and stellar models, might in fact be systematically wrong. Further work is needed, but after more than 10 years, it seems the solution to the solar abundance problem is starting to emerge. And it looks like the SSM will survive for the next battle.

This article is organized as follows. In Section\,\ref{sec:sm} we describe the basic concepts that define the SSMs, the microscopic and the macroscopic physics involved.
Section~\ref{sec:compo} gives details on the solar composition: how it enters in SSM calculations and different available compilations of solar composition. Section~\ref{sec:helio} presents general aspects of helioseismic constraints and goes to some length into the solar abundance problem, i.e. the discrepancy between SSMs with low metallicity compositions and helioseismic inferences. This section also includes some discussion on radiative opacities, highlighting the relevance of recent experimental results. Section~\ref{sec:neut} summarizes recent results in nuclear reaction rates and presents at length solar neutrino results from models and the comparison with solar neutrino experiments; it includes some results recently presented on the significance of electron capture CNO neutrinos and an overview on the current status of theoretical uncertainties. Section~\ref{sec:beyond} gives a very brief account of limitations that are intrinsic to the SSM framework, i.e. that are related to the macrophysics included in the models. Finally, Section~\ref{sec:end} contains just a few final thoughts.

\section{Solar models: setting the Standard}
\label{sec:sm}

According to the Collins English Dictionary, a standard is \emph{an accepted or approved example of something against which others are judged or measured}. The standard solar model (SSM) is, in fact, a well defined working framework for solar modeling upon which we can test our understanding of solar interior physics. But standard is also \emph{of recognized authority, competence, or excellence}; the SSM has provided, at least until recent years, an accurate description of most solar interior properties, as inferred from helioseismology and, later, by solar neutrinos.

\subsection{The SSM framework}\label{sec:ssm}

The SSM framework is defined by the physical processes, or macrophysics, included in the model, and by how the present-day solar model is computed. An important aspect of the SSM is having as few free parameters in the model as possible, and that they can be determined in a rather direct fashion from observational constraints. The goal behind being that the SSM should offer a well-defined framework that is as free from subjective choices as possible. We give some details in what follows.

\smallskip
\textbf{\emph{Initial setup.}} The SSM is the result of the evolution of an initially fully homogeneous \onemsun\ stellar model, starting from the pre-main sequence or the zero-age main sequence, up to the present-day solar system age $\tsun$. This implies the assumption that the Sun assembled its mass in a short timescale and evolved initially along the Hayashi track, where it was fully convective. Also, it implies that there has not occured appreciable mass loss afterwards.

\smallskip
\textbf{\emph{Adjustable quantities and constraints.}} The SSM is required to match, at $\tsun$, the solar luminosity $\lsun$, the solar radius $\rsun$, and the photospheric (surface) metal-to-hydrogen mass fraction $\zxsun$. The three adjustable quantities in the model are the initial helium and metal mass fractions $\yini$ and $\zini$ respectively, and the parameter $\amlt$ of the mixing length theory (or its equivalent in other prescriptions of convection). Roughly, $\amlt$ is related to $\rsun$, $\yini$ to $\lsun$ and $\zini$ to $\zxsun$, although the three adjustable quantities depend on the three observational constraints and are therefore correlated with each other. $\lsun=3.8418\times10^{33}\,\hbox{erg s}^{-1}$ and $\rsun=6.9598\times10^{10}$\,cm \cite{bahcall:2001}. $\zxsun$ is discussed below in more detail in Sect.\,\ref{sec:compo}.

\smallskip
\textbf{\emph{Physical processes.}} The physics input in the SSM is rather simple and it accounts for: convective and radiative transport of energy, chemical evolution driven by nuclear reactions, microscopic diffusion of elements which comprises different processes but among which gravitational settling dominates.

\smallskip
\textbf{\emph{Constitutive physics.}} Over the last 20 years, since the modern version of the SSM was established when microscopic diffusion was incorporated, it has been the continuous improvement of the constitutive physics what has brought about the changes and the evolution SSMs. In particular, a lot of effort has gone into experimental and theoretical work on nuclear reaction rates in this period. But changes in radiative opacities and the equation of state are also relevant. Our current choices for the SSM are as follow.
The equation of state is the 2005 version of OPAL \cite{opal:2002}, atomic radiative opacities are from the Opacity Project (OP) \cite{op}, complemented at low temperatures with molecular opacities from \cite{ferguson:2005}. Nuclear reaction rates for the pp-chains and CNO-bicycle are from the Solar Fusion II compilation \cite{sfii}. Microscopic diffusion coefficients are computed as described in \cite{thoul:1994}. Convection is treated according to the mixing length theory \cite{kipp:1990}. The atmosphere is grey and modeled according to a Krishna-Swamy $T-\tau$ relationship \cite{ks:1966}.

\section{Solar composition}\label{sec:compo}

The determination of the abundance of chemical elements in the Sun is done primarily through spectroscopy of the solar photosphere. Underlying such type of analysis are the modeling of the solar atmosphere to determine its temperature and density stratifications and detailed radiation transfer calculations that finally link elemental abundances with spectral line intensities and shapes. Mainly, the introduction of three-dimensional  radiation hydrodynamic (3D-RHD) models of the solar atmosphere and of non-local thermodynamic equilibrium calculations for line formation have led to large changes in the determinations of solar abundances. This has been discussed at length in the literature, but two relevant references summarizing the most important results, and also highlighting that no unanimous  agreement exists, are \cite{agss09,caffau:2011}. 

Table\,~\ref{tab:compo} lists the abundances determined by different authors for the most relevant metals in solar modeling: GN93\cite{gn93}, GS98\cite{gs98}, AGSS09\cite{agss09}, C11\cite{caffau:2011} and AGSS15\cite{scott:15a,scott:15b}. Abundances given in the table come, with the only exception of neon, from spectroscopic analysis of the solar photosphere. The introduction of the 3D-RHD models around year 2001 is an inflection point in the values determined for abundances of volatile elements, particularly C, N, and O. Results from the Asplund group (AGSS09), in particular, give large reductions $>30\%$ with respect to the older generation of analysis (GN93, GS98). C11 \cite{caffau:2011}, based on 3D-RHD independent models, finds CNO abundances intermediate between older generation and those of Asplund. Interestingly, three different 3D-RHD solar model atmospheres have been compared in \cite{beeck:2012}, showing minimal variations that are not the cause of the different CNO results between AGSS09 and C11. The latter must originate in the calculations of line formation, the choice of atomic data and/or the selection of lines used in the analysis.  Results for refractory elements, on the other hand, have been more robust over time. Note that from spectroscopy only abundances relative to hydrogen can be obtained because the intensity of spectroscopic lines is measured relative to a continuum that is determined by the hydrogen abundance in the solar atmosphere. The latest revision of spectroscopic results for refractories by Asplund's group, AGSS15, yields very similar results to AGSS09. A revision of the CNO abundances is underway, and small changes are expected as well.

The last row in the table gives the total photospheric present-day metal-to-hydrogen ratio $\zxsun$ and it is the quantity used as observational constraint to construct a solar model. In fact, the solar composition set used in solar models determines not only $\zxsun$ but also the relative abundances of metals in the models. In this sense, $\zini$ acts as a normalization factor that, together with $\yini$ and the relation $\xini + \yini + \zini = 1$, determines completely the initial composition of the model.
\begin{table}[!ht]
\caption{Solar photospheric composition through time and authors for most relevant metals in solar modeling. Abundances are given in the standard astronomical scale  $\log{\epsilon_i}=\log{\left(n_i/n_{\rm H}\right)}+12$, where $n_i$ is the number density of a given atomic species. \label{tab:compo}}      
\begin{tabular}{lccccc}
\hline\noalign{\smallskip}
El. & GN93 & GS98 & AGSS09 & C11 & AGSS15 \\
\noalign{\smallskip}\hline\noalign{\smallskip}
C  & 8.55  & 8.52 & 8.43 & 8.50 & --- \\
N  & 7.97  & 7.92 & 7.83 & 7.86 & --- \\
O  & 8.87  & 8.83 & 8.69 & 8.76 & --- \\
Ne & 8.08  & 8.08 & 7.93 & 8.05 & 7.93 \\
Mg & 7.58  & 7.58 & 7.60 & 7.54 & 7.59 \\
Si & 7.55  & 7.55 & 7.51 & 7.52 & 7.51 \\
S  & 7.33  & 7.33 & 7.13 & 7.16 & 7.13 \\
Fe & 7.50  & 7.50 & 7.50 & 7.52 & 7.47 \\
\noalign{\smallskip}\hline \noalign{\smallskip}
(Z/X)$_\odot$ & 0.0245 & 0.0230 & 0.0180 & 0.0209 & --- \\
\noalign{\smallskip}\hline
\end{tabular}
\end{table}

Refractory elements play a very important role in solar models. They amount to about 20\% of the total metal mass fraction and are important contributors to the radiative opacity in the solar interior, particularly Si and Fe and, to a lesser extent, Mg and S. Others like Ca, Al and Ni play minor roles and we do not consider them as relevant sources of uncertainty, although their abundances are used consistently in the calculation of radiative opacities. Importantly, abundances for refractories can be determined very precisely from chondritic CI meteorites. A detailed account and relevant references can be found in \cite{lodders:2009}. Meteoritic abundances have remained robust through the years and spectroscopic values have, over time, evolved in the direction of matching the meteoritic scales. It is therefore desirable to combine spectroscopic measurements of volatiles with the more robust meteoritic results for refractories. The solar abundance composition thus constructed is the one used in solar models discussed here, unless otherwise noticed.

As usual with the spectroscopic abundances, meteoritic ones are also relative but, in this case, the reference element is generally Si. Matching both scales then requires using an anchor point between the two. Traditionally, this is done by equating the abundance of Si between the two scales. The purely meteoritic abundances for Si and Fe are $\log{\epsilon_{\rm Si}}=7.56$ and $\log{\epsilon_{\rm Fe}}=7.50$. The photospheric Si abundance in GS98 is 7.55\,dex, and so this implies a very small -0.01\,dex shift of the whole meteoritic scale. On the other hand, for AGSS09 and its newer version AGSS15 the photospheric Si is 7.51, so the meteoritic scale has to be shifted by -0.05\,dex, about 12\% (see Table\,\ref{tab:compo}).

Interestingly, note that the new revision of solar abundances AGSS15 has a photospheric Fe abundance $\log{\epsilon_{\rm Fe}}=7.47$ compared to previous 7.50 from AGSS09. By using Si to match the meteoritic scale to AGSS15 the resulting Fe is 7.45, i.e. only 0.02\,dex is the difference between the meteoritic and the newest photospheric scale for Fe. As it has happened historically, spectroscopic abundances evolve towards meteoritic ones. This reinforces the idea that the meteoritic scale is the robust choice and the reason why rely upon it for refractory elements.

Uncertainties in element abundances are difficult to quantify. In many cases, errors quoted for a given element are just the internal dispersions of the mean obtained from abundances determined using different spectral lines. A detailed study of systematic uncertainties is not available, so uncertainties quoted in works on solar abundances can only be taken as indicative. Typical values quoted by AGSS09 are about 0.05\,dex for volatiles. Interestingly, C11 uses a different, less stringent, selection of spectral lines and finds larger values, from 0.06\,dex for C to 0.12\,dex for N. Using the meteoritic scale for refractories has the additional advantage that uncertainties are very small, typically 0.01 to 0.02\,dex and are much less prone to systematic errors as no modeling is involved.

Based on the discussion above, when possible, the meteoritic scale is used in constructing solar models. In the case of GS98 the two scales are very similar. But in the case of AGSS09 some differences are present and have some impact in solar model predictions \cite{serenelli:2009}. Therefore, we identify the combination of AGSS09 photospheric abundances for volatiles and meteoritic ones for refractories as AGSS09met in what follows. In relation to AGSS09 values given in Table~\ref{tab:compo}, AGSS09met has lower Mg and Fe by 0.07 and 0.05~dex respectively and $\zxsun=0.0178$.

\section{Helioseismology}\label{sec:helio}

For two decades now, helioseismology has provided the most stringent constraints on the interior structure of the Sun \cite{jcd:1996,bahcall:1995,gough:1996}. The measurement of the frequencies of thousands of global acoustic eigenmodes (or p-modes), with angular degrees from $\ell=0$ up to several hundred and with precisions of the order of 1 part in 10$^5$, has allowed the reconstruct the interior structure of the Sun with excellent precision. This is possible because modes with different angular degree and frequencies have different inner turning points and therefore probe regions of the Sun differentially. Moreover, low degree modes, understood here as those with $\ell=0, 1, 2, 3$ also play a very important role because they reach the innermost solar regions and help probe the solar core, where solar neutrinos are produced. 

Of particular interest for testing the quality of solar models are: the solar sound speed profile derived from inversions, the depth of the convective envelope $\rcz$, and the abundance of helium in the solar envelope $\ys$. A detailed account of the basics of helioseismology, including discussion on inversion techniques, can be found in \cite{jcd:2002,basu:2008} among many other publications. 
Also, specific combinations of frequencies of low degree modes can be used to probe the solar core, these are the so-called frequency separation ratios. They have the two advantages: they are free from uncertainties in the modeling of near surface convection, and they are dominated by the structure of the solar core \cite{roxburgh:2003,oti:2005}. The solar density profile can also be used as a  probe for solar models, but there are large correlations in the derived profiles between different parts of the Sun. They arise largely because $\msun$ is a constraint in helioseismic inversions and the density gradient very large, e.g. small differences in the solar core appear as much larger relative changes in the outer layers so that the total mass is conserved.

\subsection{Solar abundance problem}

\begin{figure}[!ht]
\resizebox{0.48\textwidth}{!}{\includegraphics{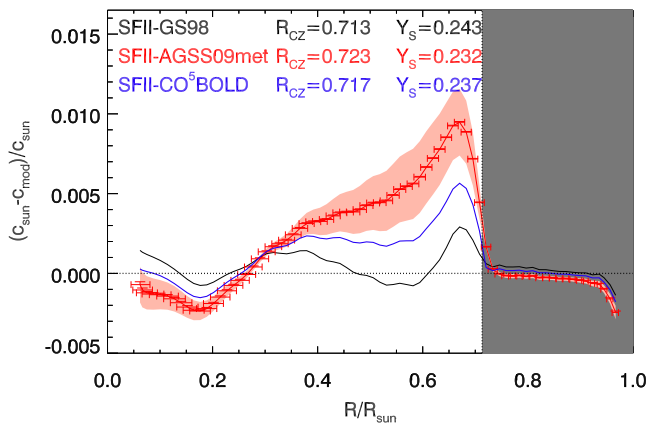}}
\caption{Relative sound speed profile and three SSMs, identified by the solar composition used in each case.\label{fig:csound}}
\end{figure}

Fig.~\ref{fig:csound} shows the relative difference in sound speed profiles for three of the sets of solar abundances discussed before. GS98 represents the older \emph{high-Z} solar abundances, and AGSS09met and CO$^5$BOLD the new 3D-RHD based solar abundances. AGSS09met in particular is representative of what we will call here the \emph{low-Z} solar abundances. The error bars depict those originating in from the seismic data and the size of the kernels used in the inversion. The shaded area represents those from solar models and guide the eye to quantify the magnitude of the discrepancy brought about by the AGSS09met composition. Model errors are to a good approximation independent of the reference model considered. Also, model results for $\rcz$ and $\ys$ are shown in the figure. Typical absolute model errors derived for these quantities from a large MonteCarlo study \cite{bahcall:2006} are, coincidentally,  0.0037 for both quantities. Newer calculations yield slightly different values; this is discussed later. These have to be compared to those obtained from helioseimic data $\rcz_{,\odot} =0.713\pm0.001$ \cite{basu:1997} and $\ys_{,\odot}=0.2485\pm 0.0034$ \cite{basu:2004}. 

The large differences in SSMs seen between high- and low-Z models have been largely discussed in the literature since 2004, generally under the name of the \emph{solar abundance problem}, tracing a parallelism with the solar neutrino problem. From a phenomenological point of view, most seismic probes, and certainly those described above, do not directly depend on the metal composition of the Sun but rather on its opacity profile. This profile is the result of atomic calculations of radiative opacities and the composition of the solar interior, and seismology is good at constraining this combination. However, it is not possible at the moment to use seismology to constrain opacities or composition separately. This has been nicely described in \cite{jcd:2009} where a low-Z SSMs has been computed with ad-hoc adjustments applied to the opacity profile such that it mimics that of a high-Z model while keeping a low-Z composition. The sound speed profile, $\rcz$ and $\ys$ from this low-Z model are practically indistinguishable from the high-Z one. 

From the above considerations, a number of ways out of the conundrum set by the low-Z abundances can be considered. The most obvious one is questioning the low-Z abundance determinations \cite{antia:2005,antia:2006,delahaye:2006,delahaye:2010}. As mentioned before, it does seem well established now that 3D-RHD models offer a good physical description of the solar photosphere and different 3D-RHD models are in good agreement with each other. Although differences between 3D-based abundances remain, e.g. AGSS09 vs C11, restoration of high-Z abundances does not seem possible on spectroscopic grounds. This statement is valid as far as models do not include drastic changes to the structure of the layers where spectral lines form. This could happen, for example, with the inclusion of chromospheres in spectral analysis, an aspect that has been almost completely overlooked so far but has been shown to impact abundance determinations of certain elements. For example, inclusion of non-local thermodynamic equilibrium and a solar chromosphere model on top of a solar atmosphere model leads to a solar Ti abundance larger by 0.14\,dex in comparison to the same model without chromosphere \cite{bergemann:2011}. It remains to be seen how this impacts the derived abundances of other elements more relevant to opacities in the solar interior such as Si, Fe and, eventually, the volatiles CNO.

Alternatively, mixing mechanisms of chemicals in the SSMs might not be correct, and additional processes could in fact affect the interior metal abundance such that it resembles a high-Z composition, while keeping abundances compatible with low-Z abundances in the surface. Increasing the rate of element diffusion (gravitational settling) or invoking accretion events early in the solar system evolution that decreased the photospheric metal abundance have been considered \cite{montalban:2004,castro:2007,guzik:2010,serenelli:2011}. Each of them offers, at best, only a partial solution to the solar abundance problem and generally accompanied by increased discrepancies in others. In particular, it has been not possible to simultaneously improve the agreement in $\rcz$ and $\ys$ \cite{delahaye:2006}.

\begin{figure}[!ht]
\resizebox{0.48\textwidth}{!}{\includegraphics{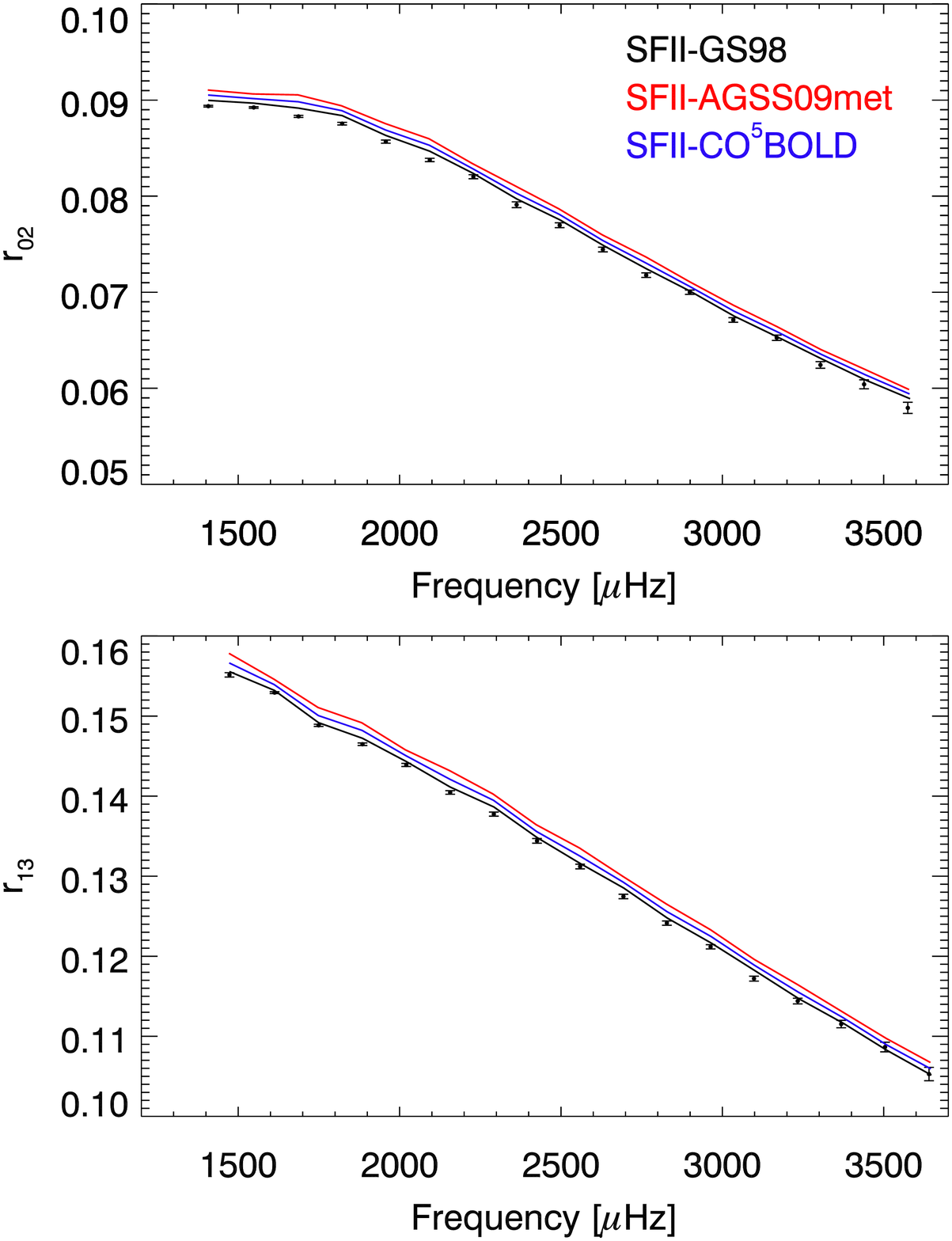}}
\caption{Frequency separation ratios as defined in Eq.\,\ref{eq:ratios}.  \label{fig:ratios}}
\end{figure}

A complementary approach to the seismic quantities discussed above can be obtained from using the frequency separation ratios. For sufficiently large radial orders and low angular degree $\ell$, the asymptotic behavior of p-modes ensures that the small frequency separations behave as 
\begin{equation}
\nu_{n,\ell} - \nu_{n-1,\ell+2} \approx -(4\ell+6) \frac{1}{4\pi^2 \nu_{n,\ell}}\int_0^{R_\odot}{\frac{dc_s}{dr}\frac{dr}{r}},
\end{equation}
where $n$ denotes the radial order ($n \gg 1$) and $\ell=0, 1$. The integrand is sensitive to the gradient of the sound speed in the central solar regions. The condition that $\ell$ is small ensures that modes are actually probing that region. It has been shown \cite{roxburgh:2003} that it is even more useful to consider the frequency separation ratios to highlight core conditions in the Sun
\begin{equation}
r_{0,2} = \frac{\nu_{n,0} - \nu_{n-1,2}}{\nu_{n,1}-\nu_{n-1,1}} \ \ {\rm and} \ \ 
r_{1,3} = \frac{\nu_{n,1} - \nu_{n-1,3}}{\nu_{n+1,0}-\nu_{n,0}}. \label{eq:ratios}
\end{equation}
In \cite{basu:2007,chaplin:2007} these frequency ratios have also been used to test SSMs extensively in the light of the solar abundance problem. The Monte Carlo set of SSMs used in \cite{bahcall:2006} was also used to construct the distribution functions for $r_{0,2}$ and $r_{1,3}$ for high- and low-Z models. Both works show again that seismic probes are not consistent with current state-of-the-art low-Z SSMs. It should be noticed, however, that $r_{0,2}$ and $r_{1,3}$ are also sensitive to the solar metallicity only through the radiative opacity. A comparison between GS98 and AGSS09 models has been presented in  \cite{serenelli:2009}. The frequency separation ratios of the same three models from Fig.\,\ref{fig:csound} are shown in Figure\,\ref{fig:ratios} and compared to solar results derived from 4752 days of observation by the BiSON experiment \cite{chaplin:2007}. Error bars from helioseismology are also shown, but they are minuscule. 

Helioseismology can potentially yield a determination of the metallicity of the Sun that is almost independent of opacities. All relevant quantities describing acoustic modes in the adiabatic approximations are comprised in the relation
\begin{equation}
c_s^2= \Gamma_1\frac{p}{\rho}, \ \ \ \rm{with}  \ \ \ \
\Gamma_1= \left(\frac{\partial \log{p}}{\partial \log{\rho}}  \right)_{\rm ad}
\end{equation}
where $c_s$ is the adiabatic sound speed, $p$ the pressure, $\rho$ the density, and
$\Gamma_1$ is the adiabatic index that equals 5/3 for an ideal completely ionized gas. A rapid spatial variation in any of these quantities leaves an imprint in the oscillation frequencies of p-modes due to partial reflection of the waves. The partial ionization of helium, particularly HeII, produces a $\sim 0.1$ depression in $\Gamma_1$ around $R \sim 0.98\,\rsun$ that has been used to determine the helium abundance in the solar photosphere (see e.g. \cite{basu:2004}). However, partial ionization of metals also produce a depression in $\Gamma_1$, albeit of a much smaller magnitude. Using seismic inversions to obtain the solar $\Gamma_1$ profile, if the signal from HeII can be extracted, then the residuals are due to a combination of different partial ionization stages of different metals. This is a subtle signal (less than one part per thousand) that extends down to $R \sim 0.70\,\rsun$, and is dominated at ever deeper regions by CV, NVI, OVII, NeIX (see in particular Fig.\,8 in \cite{basu:2008}). The idea has been used \cite{antia:2006,lin:2007} to determine $Z_\odot=0.0172 \pm 0.002$, i.e. $\zxsun \sim 0.0234$, in agreement with high-Z abundances. However, this result has recently been challenged by \cite{vorontsov:2013} who, using a different analysis method of seismic data and SAHA-S3, a different equation of state \cite{baturin:2013}, have found the much lower range $Z_\odot=0.008-0.013$ depending on the datasets and seismic techniques used. The low end of the range is simply too low to be believable as it would require enormous changes in all solar modeling: interior, atmospheric and spectroscopic. The high end of the range, on the other hand, is consistent with the AGSS09 composition. The subtlety of the $\Gamma_1$ signal left by metals and the systematic differences depending on the techniques employed make this type of measurement very delicate and, currently, highly uncertain. Moreover, transforming such measurement into a metal abundance requires a equation of state which, however, can hardly be tested independently to such accuracy. While determining the solar metallicity from $\Gamma_1$ is a tantalizing idea, the robustness of the technique remains to be proven.

We close the section devoted to helioseismology with a brief comment on g-modes. These modes, for which buoyancy is the restoring force, have a propagation cavity in the Sun different than the p-modes and are damped in the solar convective envelope. The expected amplitudes are extremely small rendering their detectability very difficult. Potentially, they are very interesting probes of the conditions in the solar core because they reach the center, unlike all p-modes with $\ell \ge 1$. In 2007, a claim that g-modes had been finally detected in the GOLF \cite{golf} data was published \cite{garcia:2007}. To date, this is a controversial result. We end the section by quoting an extract from the abstract in the review paper on g-modes \cite{appor:2010}: \emph{``The review ends by concluding that, at the time of writing, there is indeed a consensus amongst the authors that there is currently no undisputed detection of solar g modes.''}

\subsection{Radiative opacity in the solar interior}

As just discussed, all robust helioseismic probes of the solar interior are sensitive to the radiative opacity profile. In the context of the solar abundance problem, \cite{jcd:2009} have shown an almost perfect degeneracy between solar composition and opacities for standard seismic probes. \cite{villante:2010}, using a different approach based on so-called Linear Solar Models (LSM) has also shown that helioseismic data and solar neutrinos put constraints on the solar opacity profile. In fact, changes of the order of 15-20\% in the radiative opacity are required to bring back the agreement between helioseismology and SSMs if the low-Z solar composition is  adopted. This differences smoothly decrease inwards to a few percent in the solar core region. The same conclusion has been reached \cite{villante:2014} when uncertainties in SSMs are completely accounted for and a global analysis combining solar neutrinos and helioseismic data is performed: there is no freedom in SSMs that can compensate the impact of reducing $\zxsun$ other than increasing the opacity. 

Two sets of radiative atomic opacity calculations have been widely used in solar model calculations, OPAL \cite{iglesias:1996} and OP \cite{op}. Differences between OPAL and OP Rosseland mean opacities across the solar radiative interior are never larger than 3\% \cite{op}, much smaller than needed for compensating changes in solar abundances. More recently, a new set of opacities has been presented \cite{opas}. An element by element comparison with OP shows, interestingly, large differences that can reach 40\% at conditions similar to the base of the convective envelope. However, OPAS yields lower opacities for intermediate Z elements such as Ne, Mg, Si and higher for high-Z elements such as Fe and Ni. Also, as discussed in \cite{iglesias:2015}, changes in opacities for a given element do not add linearly to the Rosseland mean.  When the Rosseland mean is compared, OPAS agrees with OPAL and OP within 4\% in the whole solar radiative interior. 

Does the agreement between different radiative atomic opacity calculations to within a few percent mean they are accurate to this level? Experimental determination of opacities for solar conditions is extremely challenging because of the combination of high temperatures and densities. The less extreme conditions this kind of experiments need to reach to be spot on the right conditions in solar interior are those at the base of the convective envelope, where T$\approx 2.35\times 10^6$\,K and $\rho\approx0.2$\,g\,cm$^{-3}$ ($n_e \approx 1\times10^{23}$\,cm$^{-3}$). It has not yet been possible to reach these conditions experimentally, but recent results are quite close \cite{bailey:2015}.

Fe is a very important contributor to solar opacities. At the base of the convective envelope it contributes about 25\% of the total opacity \cite{basu:2008} and therefore has a preponderant impact on seismic properties of SSMs \cite{villante:2014}. This is because Fe is both abundant in the Sun and with a complex atomic structure. For the latter reason, it is also challenging for current atomic models. Very recently, results from experiments carried out at the Z-facility at Sandia labs \cite{bailey:2015} have measured the wavelength (monochromatic) dependent Fe opacity. Experiments at four different thermodynamic conditions have been performed, with the (T,$n_e$) values (1.91,0.07), (1.97,0.2), (2.11,0.31), and (2.26,0.4) where T is in $10^6$\,K and $n_e$ in $10^{23}$\,cm$^{-3}$. In all conditions under which experiments were carried out, results show very large differences with all available opacity calculations. The Rosseland mean opacity for Fe determined from experiment is on average 60-65\% larger than any predicted value from atomic calculations. When experimental results for Fe are combined with OP theoretical calculations for all other elements (there is no experimental data for any other element at these conditions), the final Rosseland mean is 7\% larger than the OP value used in SSMs discussed here. It has to be kept in mind that conditions reached at the Z-facility are not yet those prevailing at the base of the CZ. The electron density is still lower by approximately a factor of 2.5. However, there seems to be clear indication that radiative opacities could be underestimated in atomic calculations by a fraction much larger than differences between different theoretical calculations would suggest. The 7\% increase in Fe opacity found by \cite{bailey:2015} would be enough, if extrapolated to solar conditions, to provide between 1/3 and 1/2 of the missing opacity when the AGSS09met composition is used in SSMs. Interestingly, this increase in opacity would practically restore the agreement between SSMs and helioseismology if the C11 composition is adopted. A direct extrapolation is not, however, physically correct and atomic models are required for modeling it on a physically sound basis.

\section{Solar neutrinos}\label{sec:neut}

The original motivation for computing ever more accurate and precise SSMs was the solar neutrino problem, dating back to almost 50 years now. An account of historic developments of solar neutrino experiments and the solar abundance problem is presented elsewhere \cite{haxton:2013}. Here we limit the presentation to the state of the art of SSMs solar neutrino results\footnote{A full tabulation of SSMs with GS98 and AGSS09 solar compositions is available at \url{http://www.ice.csic.es/personal/aldos/Solar_Models.html}}. 

\subsection{Astrophysical factors}

The accurate and precise calculation of solar neutrino fluxes from SSMs require excellent quality data for many of the nuclear reactions of the pp-chains and the CNO-bicycle that control hydrogen fusion in the solar interior. Much work has been done in this respect, even continuing after the solution of the solar neutrino problem. Here, we cannot make a full review of them and the reader is referred to the Solar Fusion II article \cite{sfii} for a full account of the experimental and theoretical developments in the preceding decade. Of particular relevance, however, have been the efforts at Gran Sasso by LUNA (Laboratory for Underground Nuclear Astrophysics, \cite{luna}) to reach very low energies.  ${\rm ^2H(p,\gamma)^3He}$ and ${\rm ^3He(^3He,2p)^4He}$ have indeed been measured at the Gamow peak and, although the first one is not crucial for solar models, the second one plays an important role in the bifurcation between pp-I and the other two pp-chains. Above the Gamow peak for solar conditions, ${\rm ^3He(^4He,\gamma)^7Be}$, another fundamental reaction for the $^7$Be and $^8$B solar neutrinos has also been measured by LUNA and, largely to these efforts, is now determined to just a few percent precision. Finally, LUNA measurement of the ${\rm ^{14}N(p,\gamma)^{15}O}$ rate has had a strong impact in stellar astrophysics in general. The astrophysical factor\footnote{Nuclear reaction cross section are usually written as $\sigma(E) = \frac{{\rm S}(E)}{E} \exp{^{-2\pi \eta(E)}}$, where ${\rm S}(E)$ is the astrophysical factor and $\eta(E)$ the Sommerfeld parameter. The leading order of a series expansion, ${\rm S}(0)$, is generally a good approximation of ${\rm S}(E)$ at solar conditions, although higher orders are used when ${\rm S}'(0)$ and ${\rm S}''(0)$ can be determined either from experiments or theory.}  obtained by LUNA is about a factor of 2 lower than previously used \cite{sfi,nacre} and, although this value had been anticipated based on older data but different extrapolation methods \cite{descouvemont:2002}, LUNA results have settled this matter. Unfortunately, this reduction translates directly into the same reduction of solar model predictions for CN fluxes. After the SFII work, the ${\rm p(p,\nu_ee+)^2H}$ rate has been revised, based on chiral effective field theory calculations. Results are consistent with previous values \cite{sfii}, although the quoted error is only 0.15\%, almost an order of magnitude smaller. 

\begin{table}
\caption{Astrophysical factors for most relevant reactions from SFII and SFI compilations. Errors are given in brackets. \label{tab:rates}}
\begin{tabular}{lcc}
\hline \noalign{\smallskip}
Reaction & SFII & SFI \\
& (keV-b) & (keV-b) \\
\hline 
${\rm S}_{11}$ & $4.01\times10^{-22}\,[1\%]$ & $3.94\times 10^{-22}\, [0.4\%]$\\
${\rm S}_{33}$ & $5.21\times 10^3\,[5.2\%]$ & $5.4\times 10^3\, [6\%]$\\
${\rm S}_{34}$ & $0.56\,[5.4\%]$ & $0.567\, [3\%]$ \\
${\rm S}_{17}$ & $2.08\times 10^{-2}\,[7.7\%]$& $2.14\times 10^{-2}\, [3.8\%]$\\
${\rm S}_{1,14}$ & $1.66 \,[7.5\%]$ & $1.57\, [8\%]$ \\
\noalign{\smallskip}\hline \noalign{\smallskip}
R(pep)/R(pp) & $\uparrow$ 2.5\% & --- \\
\noalign{\smallskip}\hline
\end{tabular}
\end{table}

Solar model results for neutrinos presented here are based on the SFII recommended rates. This choice is done because SFII critically accounts for different sources of data for each reaction, even at the expense of recommending larger errors than the original individual sources for certain reactions. A summary of the most relevant changes with respect to the Solar Fusion I rates \cite{sfi} is given in Table\,\ref{tab:rates}. Other rates not listed have lesser of an impact on the solar neutrino fluxes and uncertainties.

\begin{figure}[!ht]
\resizebox{0.48\textwidth}{!}{\includegraphics{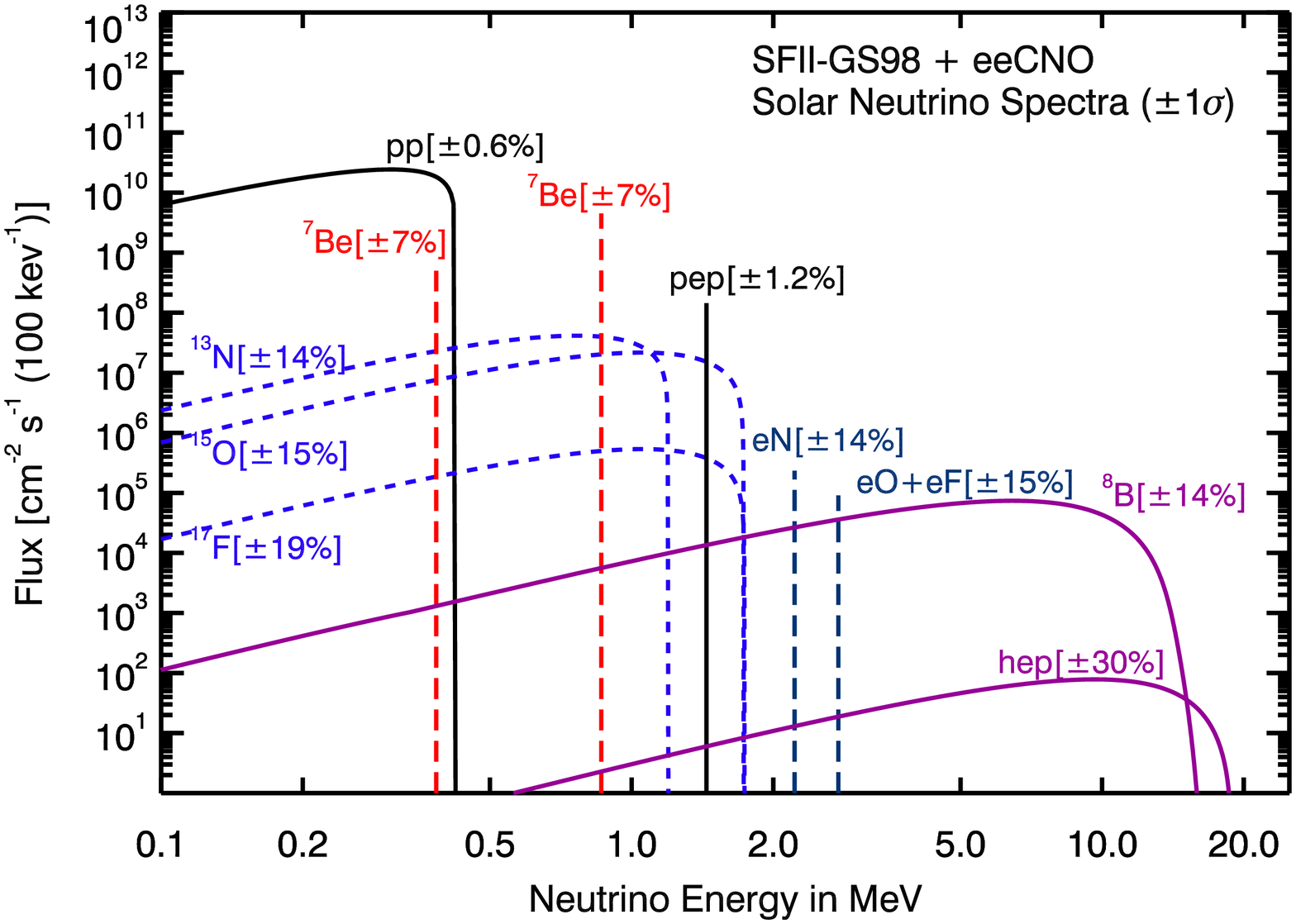}}
\caption{Spectrum of solar neutrino fluxes corresponding to the SFII-GS98 model. ecCNO neutrinos \cite{eecno} have been added in addition to standard fluxes. Electron capture fluxes are given in $\hbox{cm$^{-2}$s$^{-1}$}$. \label{fig:nuspec}}
\end{figure}

Table~\ref{tab:neut} lists the predicted solar neutrino fluxes for the eight reactions associated with the pp-chains and CNO-bicycle for the three solar models discussed in the previous section. Additionally, electron capture CNO neutrinos (ecCNO) are given \cite{eecno}. Errors are given in brackets. Note that with respect to the original publication \cite{serenelli:2011}, the error in $^{17}$F is now larger because of including the 7.5\% contribution from the $^{16}{\rm O(p,\gamma)^{17}F}$ astrophysical factor S$_{116}$ \cite{sfii}. 
Figure~\ref{fig:nuspec} shows the solar neutrino spectrum from \cite{serenelli:2011} for the SFII-GS98 SSM and the ecCNO neutrinos from \cite{eecno}.

\subsection{Solar neutrinos from pp-chains}

Differences in the fluxes associated to the pp-chains are exclusively due to changes in the core temperature in the models. Considering central temperatures as indicative of the changes, these are 1.562, 1.553, and 1.546$\times10^7$K for the GS98, C11 and AGSS09met models respectively, i.e. differences that are 1\% at most. Differences in the core temperature arise because of the changes in the radiative opacity that is lower for lower metal abundances in the solar interior. The radiative opacity determines the radiative gradient in the solar interior, so a lower opacity leads to an overall smaller core temperature, as just described. It is interesting to note, therefore, that analogously to what has been described in the previous section, pp-chain neutrino fluxes are not directly sensitive to the solar composition, but to the radiative opacity. Therefore, a degeneracy is present as the latter depends both on the composition and on theoretical models of radiative opacity calculations.

\begin{table}
\caption{Neutrino fluxes. Columns two to four give results from different SSMs identified according to the solar abundance used, last column solar neutrino fluxes determined from neutrino experimental data and the luminosity constraint, but independently of SSMs. Errors are quoted in brackets. Last row: agreement between SSMs and solar neutrino fluxes. Units are, in 
cm$^{-2}$\,s$^{-1}$: $10^{10}$ (pp), $10^{9}$ ($^7$Be), $10^8$ (pep, $^{13}$N, $^{15}$O), $10^6$ ($^8$B, $^{17}$F), $10^5$ (eN, eO) and $10^3$ (hep, eF). \label{tab:neut}}
\begin{tabular}{lcccc}
\hline \noalign{\smallskip}
 & SFII & SFII & SFII &\\
Flux & GS98 & C11 & AGSS09met &  Solar\\
\hline
pp & $5.98\,[0.6\%]$ & 6.01  & 6.03  & $6.05\,[0.6\%]$ \\ 
pep & $1.44\,[1.1\%]$ & 1.46 & 1.47  & $1.46\,[1.2\%]$ \\  
hep &  $8.04\,[3\%]$ & 8.19 & 8.31 &$ 18\,[45\%]$\\ 
$^7$Be & $5.00\,[7\%]$ & 4.74 & 4.56 & $4.82\,[4.5\%]$\\ 
$^8$B  & $5.58\,[14\%]$ & 4.98 & 4.59 & $5.00\,[3\%]$ \\ 
$^{13}$N & $2.96\,[14\%]$ & 2.62 & 2.17 & $\leq 6.7 $ \\ 
$^{15}$O & $2.23\,[15\%]$ & 1.92 & 1.56 & $\leq 3.2 $\\ 
$^{17}$F & $5.52\,[19\%]$ & 4.27 & 3.40 & $\leq 59 $\\
\noalign{\smallskip}\hline \noalign{\smallskip}
$\chi^2/P^{\rm a}$ & 3.5/90\% & 3.2/92\% & 3.4/90\% & --- \\
\noalign{\smallskip}\hline \noalign{\smallskip}
eN & $2.34\,[14\%]$ & 2.07 & 1.71 & --- \\ 
eO & $0.88\,[15\%]$ & 0.76 & 0.62 & --- \\ 
eF & $3.24\,[19\%]$ & 2.51 & 2.00 & --- \\
\noalign{\smallskip}\hline \noalign{\smallskip}
\end{tabular}
\end{table}

Solar neutrino fluxes linked to the pp-chains can be determined from neutrino experiments without recourse to solar models. The $^8$B and $^7$Be neutrinos are particularly well determined. For $^8$B, results from the different phases of SNO \cite{SNOI,SNOII,SNOIII,SNOIV} and SuperKamiokande \cite{superkI,superkII,superkIII} lead to an experimental determination of only 3\%. For $^7$Be, Borexino results \cite{borexino:be7_1,borexino:be7_2,borexino:review} allow a determination with only a 4.5\% uncertainty. A global analysis \cite{bahcall:2003} of neutrino experiments allows in turn the determination of the pp and pep fluxes. Results of such analysis \cite{serenelli:2011} are given in the last column of Table~\ref{tab:neut}. The small uncertainty in the pp and pep fluxes comes from using the luminosity constraint in the analysis. 

When the solar luminosity constraint is not imposed in the solar neutrino analysis, the pp and pep fluxes are much more unconstrained. In fact, such analysis, but using only the initial 192 days of Borexino data that determined $^7$Be with $\sim 10\%$ uncertainty, shows  that without the luminosity constraint the pp and pep fluxes could be constrained to $\sim 15\%$ \cite{concha:2010}. This uncertainty could be partly reduced by using the final Borexino Phase I results that narrow down the $^7$Be uncertainty to just $\sim 4.5\%$, but the uncertainty in the pp flux derived in this way still remains too high to allow very meaningful tests on solar energetics. 

A more stringent test of the origin of the solar luminosity therefore relies on measuring the pp flux or its close relative pep. In fact, Borexino has provided the first direct evidence for the pep neutrinos \cite{borexino:pepcno,borexino:review} with an interaction rate R(pep)= $3.1 \pm 0.6_{\rm stat} \pm 0.3_{\rm syst}$ cpd/100 ton. In the 3-flavor neutrino oscillation framework this translates into a $1.63 (1 \pm 0.20) \times 10^8$\,cm$^2$\,s$^{-1}$ pep flux, well in agreement with SSMs albeit with a still large uncertainty. 

More recently, the formidable task of measuring the pp flux has been achieved by Borexino \cite{borexino:pp}. The flux, $6.6(1 \pm 0.11) \times 10^{10} \hbox{cm$^{-2}$s$^{-1}$}$, is nicely consistent with solar models that predict $\sim 6.0\times 10^{10} \hbox{cm$^{-2}$s$^{-1}$}$ for this flux. Based on solar models, it is expected that the pp-chains provide $\sim 99\%$ of the total nuclear energy. Because the ${\rm p(p,e^-\nu_e)^2H}$ reaction initiates and regulates the pp-chains, the Borexino measurement by itself establishes that $\lsun = 1.1 (1 \pm 0.1) L_{\rm pp}$, where $L_{\rm pp}$ is the total power generated by pp-chains. 

\begin{figure}[!ht]
\resizebox{0.45\textwidth}{!}{\includegraphics{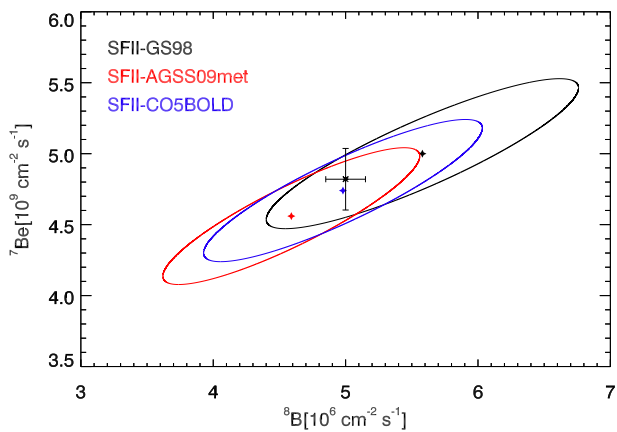}}
\caption{Experimental and model results for the $^7$Be and $^8$B solar neutrino fluxes. Experimental results are derived using all neutrino data, but are dominated by SNO, SuperKamiokande and Borexino results. Colored stars denote the central values for each SSM. \label{fig:be7b8}}
\end{figure}

The last row in Table~\ref{tab:neut} shows the $\chi^2$ obtained from comparing solar models to solar neutrino fluxes. All three models are in equal agreement with the experimental results. The luminosity constraint has been used in deriving the solar fluxes. Then, the agreement between solar models and the $^7$Be and $^8$B experimental results directly imply agreement for the pp and pep fluxes. The variations of pp-chain fluxes for models with different compositions are very easily understood with the temperature dependence of solar fluxes \cite{bahcall:1996} and the differences found in the core temperatures given above. Figure~\ref{fig:be7b8} shows the comparison between the experimental and model results for $^7$Be and $^8$B fluxes, including the strong correlation present in models. The displacement of the theoretical results in this plane occur along the major axis of the correlation between the $^7$Be and $^8$B fluxes, which is determined by their relative temperature dependences. This graphically shows that the solar composition only impacts these fluxes indirectly, by altering the solar core temperature profile.

\subsection{Solar neutrinos from CNO-bicycle}

Solar models and solar neutrino experiments show that the CNO-bicycle is a marginal source of energy in the Sun, with only 1\% of the solar energy being produced by it, particularly by the CN-cycle. The immediate implication is that the expected CN fluxes are low (we refer here as CN fluxes to $^{13}$N and $^{15}$O and drop $^{17}$F altogether from the discussion). The advantage, however, is that the physical conditions in the solar core, particularly temperature profile, are not established by the CN-cycle but by the pp-chains. Therefore, the CN-cycle retains a linear dependence on the C+N abundance in the solar core that cannot washed out by temperature variations. CN neutrino fluxes can be used to determine solar core abundances (at least the added C and N) provided other sources of uncertainties in models are under control. 

\begin{figure}[!ht]
\resizebox{0.45\textwidth}{!}{\includegraphics{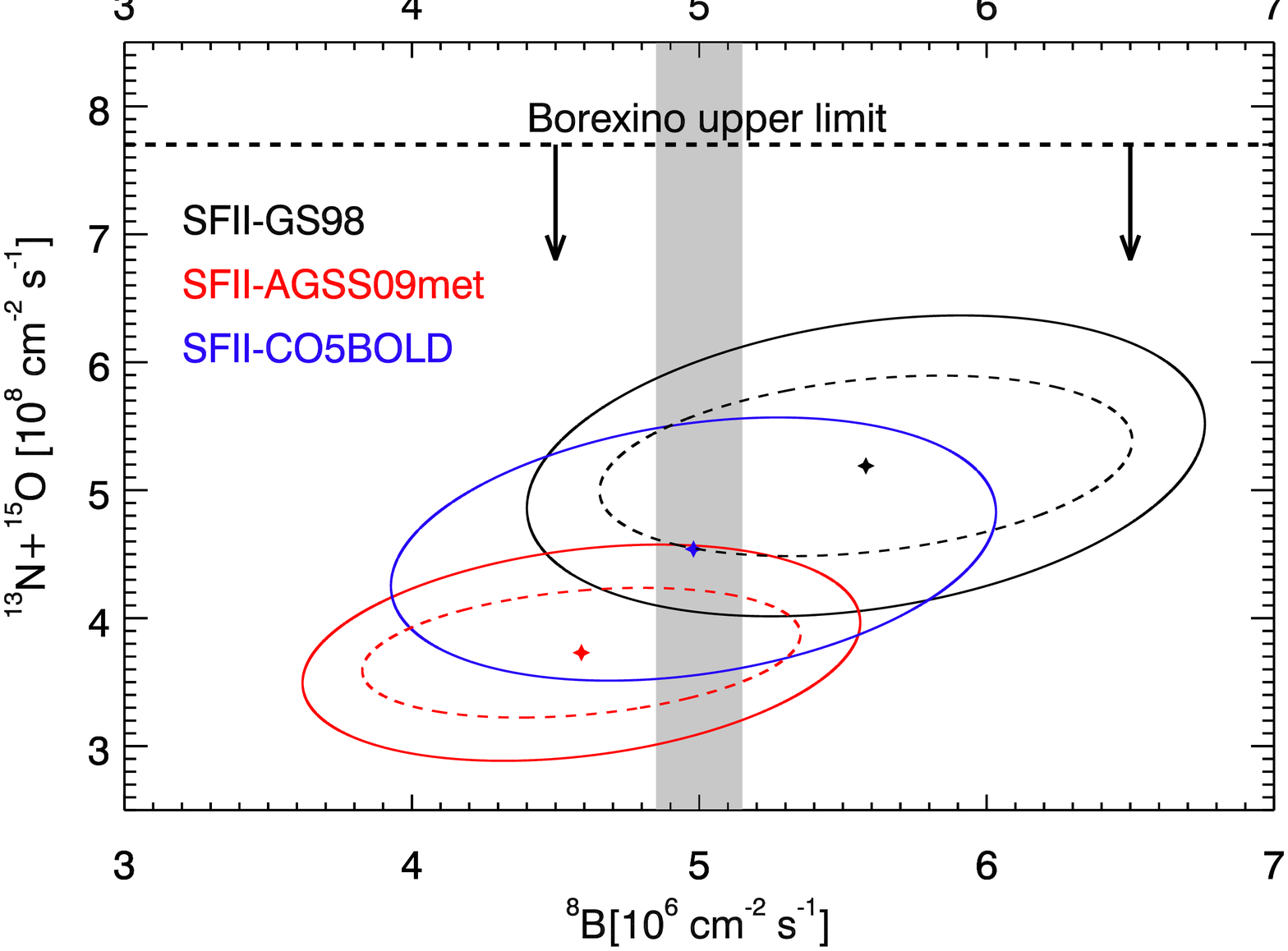}}
\caption{Experimental and model results for the added CN solar neutrino fluxes. 
Solid lines: results including all sources of uncertainty. Dashed lines: excluding solar composition uncertainty. The upper limit from Borexino is still higher than SSM results regardless of the solar composition adopted.
\label{fig:cnob8}}
\end{figure}

The best direct limit to the CN fluxes comes from Borexino \cite{borexino:pepcno,borexino:review}, where an upper limit of $7.7\times10^8$\,cm$^2$\,s$^{-1}$  was determined. Indirect constraints obtained from global fits to solar neutrino data that do not include the Borexino limit yield somewhat larger upped bounds,  $9.9\times10^8$\,cm$^2$\,s$^{-1}$ \cite{serenelli:2011}, as shown in Table~\ref{tab:neut}. It is still about 40\% higher than the GS98 SSM model prediction and so it is not possible at the moment to play a role in discriminating different solar abundance results. A comparison of the added CN fluxes against $^8$B is shown in Figure~\ref{fig:cnob8}, where the current Borexino limit is also shown. Solid lines show predictions including all sources of errors are accounted for as shown in Table~\ref{tab:neut}, including those from C and N abundances, which amount to 12\%. The dashed lines show the same results for SFII-GS98 and SFII-AGSS09met but without including the solar composition uncertainty. The difference in central values of the added fluxes between the two extreme models, SFII-GS98 and SFII-AGSS09met, is 39\% computed with respect to the SFII-AGSS09met. Unlike the case of $^7$Be and $^8$B, here displacement of model results are not aligned with the correlation between the two fluxes; this is due to the additional dependence of CN fluxes on C+N abundance, not related to temperature variations.

The discriminating power of CN fluxes can be enhanced and made more model independent by separating different classes of sources of errors in the models and by using other experimental data as probes of actual solar core conditions. In particular \cite{haxton:2008,serenelli:2013} have shown that the exquisite precision of current $^8$B experimental result and the $^8$B large temperature sensitivity can be used as an efficient thermometer that isolates the so-called \emph{environmental} uncertainties, i.e. those that impact the temperature profile, from those that affect differentially individual solar neutrino fluxes. Among the latter there are the astrophysical factors (S$_{\rm 17}$, S$_{\rm 114}$) and the C and N abundances. In particular, a relation between the $^{15}$O and $^8$B fluxes can be established \cite{serenelli:2013} such that 
\begin{eqnarray}
\frac{{\rm \Phi(^{15}O)}}{{\rm \Phi(^{15}O)^{SSM}}} \Big/ \left[\frac{{\rm \Phi(^8B)}}{{\rm \Phi(^8B)^{SSM}}}\right]^{0.785} \Big. = \left[ \frac{C+N}{C_{\rm SSM} + N_{\rm SSM}} \right] \times \nonumber \\ \big[1 \pm 2.6\% ({\rm Diff}) \pm 10.6\% ({\rm nuc}) \big]. \ \ \ \ \ \ \ 
\label{eq:b8cno}
\end{eqnarray}
The first error term is all that remains of the uncertainties from environmental factors ($\lsun$, $\tsun$, opacity, elements heavier than N) and it is due to the settling of heavy elements in the solar core that affects CN fluxes not just by altering the core temperature but also by increasing the CN abundance that catalyzes the CN-cycle. The second error source is from astrophysical factors and it is equally dominated by S$_{\rm 17}$ and S$_{\rm 114}$. 
The current 3\% uncertainty in the solar $^8$B flux makes it a small error source. Therefore, given a measure of $^{15}$O, the C+N abundance can be extracted almost directly from the relation  above. To be noted, SSM plays a role only as a scaling factor and in the determination of the exponents linking changes in the input parameters to the solar temperature. These dependences are robust to variations in the SSM, so the expression above is virtually independent of SSMs. 
A similar expression can be derived for $^{13}$N or any combination of the two CN fluxes appropriate for the characteristics of a neutrino detector.

Equation\,\ref{eq:b8cno} above establishes a relation from which it is possible to determine the solar composition with an intrinsic error of 11\% that could be further reduced by improvements on uncertainties in the  S$_{17}$ and S$_{114}$ astrophysical factors, that completely dominate the error budget. A measurement of $^{15}$O with 10\% uncertainty would allow a determination of the core C+N abundance with $\sim15\%$, an uncertainty comparable to the best spectroscopic measurements and more than a 2$\sigma$ result in terms of differences between SFII-GS98 and SFII-AGSS09met solar models.

\subsection{Solar neutrinos from non-standard sources}

The expectation values for solar neutrino fluxes from reactions that are not part of the standard CNO-bicycle have been recently reanalyzed \cite{eecno}. In particular, the so-called ecCNO neutrinos from $e^-$ captures on the unstable nuclei $^{13}$N (eN), $^{15}$ (eO) and $^{17}$F (eF) have been considered. Their fluxes are largely determined by the branching between $e^-$ capture and $\beta^+$ decay rates. Absolute values are small, typically three orders of magnitude lower than their $\beta^+$ counterparts. On the other hand, these fluxes have the interesting property they are monochromatic. The energies of the e$^{13}$N, e$^{15}$O and e$^{17}$F fluxes are 2.22,  2.754 and 2.761~MeV respectively and they probe the interesting energy transition region between vacuum and matter enhanced neutrino oscillations. Interestingly, \cite{eecno} concludes the ecCNO fluxes cannot be ignored when reconstructing the low-energy upturn of the survival probability of electron neutrinos. Also, the possibility of detecting these neutrinos with next generation of neutrino experiments, including a thorough discussion on the role of backgrounds, is discussed in that reference.

\subsection{Theoretical uncertainties}

Currently, the error budget of all solar neutrinos is the result of adding several comparable contributions. The only exception is the hep flux that is completely dominated by the uncertainty of  the $^3{\rm He(p,\nu_e)^4He}$ rate. Also, the added C+N abundance dominate uncertainties for CN fluxes. Table~\ref{tab:errors} gives a ranking of the five more important uncertainty sources for each neutrino flux and its magnitude. 

\begin{table}
\caption{Dominant sources of theoretical errors for solar neutrinos fluxes. Symbols have their usual meaning, except for $\kappa$ and D that represent radiative opacity and microscopic diffusion respectively.
\label{tab:errors}}
\begin{tabular}{lrrrrrr}
\hline \noalign{\smallskip}
pp &       $\lsun$:0.3\% & $\kappa$:0.3\% & S$_{34}$:0.3\% & D:0.2\% & S$_{33}$:0.2\% \\ 
pep &      $\kappa$:0.8\% & S$_{34}$:0.5\% & $\lsun$:0.4\% & D:0.3\% & S$_{33}$:0.2\% \\  
hep &      hep:30\% & S$_{33}$:2.3\% & D:0.5\% & S$_{34}$:0.4\% & Ne:0.3\% \\ 
$^7$Be &   S$_{34}$:4.6\% & $\kappa$:3.0\% & S$_{33}$:2.1\% & D:1.8\% & $\lsun$:1.4\% \\ 
$^8$B  &   S$_{17}$:7.7\% & $\kappa$:6.7\% & S$_{34}$:4.3\% & D:3.8\% & $\lsun$:2.8\%\\ 
$^{13}$N & C:11\% & S$_{114}$:5.4\% & D:4.8\% & $\kappa$:3.7\% & S$_{11}$:2.1\% \\ 
$^{15}$O & C:9.8\% & S$_{114}$:7.2\% & D:5.5\% & $\kappa$:5.2\% & S$_{11}$:2.9\%\\ 
$^{17}$F & O:13.3\% & S$_{116}$:7.5\% & $\kappa$:5.8\% & S$_{11}$:3.1\% & $\lsun$:2.6\% \\
\noalign{\smallskip}\hline \noalign{\smallskip}
\end{tabular}
\end{table}

Astrophysical factors S$_{17}$ and S$_{114}$ are the two most critical contributions to current model uncertainties (S$_{116}$ is comparable, but prospect for measuring the $^{17}$F flux remain remote). They are even more important if an analysis based on Eq.\,\ref{eq:b8cno} is performed because in that case they dominate the total error budget. From stellar physics, radiative opacities and, to a lesser extent microscopic diffusion play a relevant role. Consistent determinations of opacity uncertainties are not available.. Here, uncertainties from opacities are obtained by correcting by a multiplicative factor (2.5\%) to the entire opacity profile.  This has been shown to underestimate variations in the sound speed profile \cite{villante:2010}, but it is a better justified approach for solar neutrinos that originate in relatively small region of the Sun.

\section{Beyond the SSM}\label{sec:beyond}

Non-SSMs are not the topic of this article, so this is a brief section and just peeks under the surface. Some of these topics have been reviewed recently \cite{turck:2011,gough:2015}.

There are well known inherent limitations in SSMs because of the underlying assumptions made. Most obviously, the Sun is a three dimensional star and spherical symmetry is good approximation only in a global sense. Dynamical processes in the Sun are intrinsically multi-dimensional. Convection and overshooting, rotation and the associated transport of chemicals and angular momentum, magnetic field dynamics are all processes that cannot be modeled in spherical symmetry. 

Among the well-known problems with SSMs is the lithium depletion observed in the Sun, that is today more than a factor of a hundred lower in the solar photosphere than in meteorites \cite{agss09}. Phenomenologically, this can be explained by some extra mixing below the convective envelope \cite{schlattl:1999,andrassy:2013}. A similar solution, perhaps including a smoother transition between the adiabatic and radiative temperature gradients, can be applied to the problem posed by the gradient of the mean molecular weight profile at the base of the convective envelope that, in SSMs, is steeper than helioseismic inversions show \cite{antia:1998,jcd:2011}. In fact, \cite{delahaye:2006,villante:2014} find that helioseismic data prefers a reduced efficiency of microscopic diffusion, consistent with the need of extra mixing around the tachocline region.

Additional limitations with SSMs are those related to rotation. The internal solar rotation profile is known with great precision from almost the surface down to 0.2\,$\rsun$ \cite{thompson:2003,eff:2008} and, in the radiative interior is consistent with rigid body rotation, i.e. a flat rotation profile. Clearly, by construction SSMs are non-rotating models. Solar models including rotation have been computed for many years, initially trying to reproduce the solar surface rotation rate of 2\,km\,s$^{-1}$ \cite{pinso:1989}. Current generation of 1D solar rotating models that account for angular momentum transfer in the solar interior and losses through magnetic winds \cite{palacios:2006,turck:2010} are also calibrated to reproduce the surface rotation rate, but can also be tested against the internal rotation profile. Results are, however, in stark disagreement with the data. Models fail to reproduce the flat rotation profile and also predict much higher (by a factor of up to 20) inner angular velocity. Transport of angular momentum is very poorly understood, if at all, in current solar and stellar modeling. 

Modeling dynamical effects in solar (and stellar) models from first principle physics is unlikely to provide an accurate physical picture of the Sun's interior. It seems unavoidable that one has to rely on sophisticated multi-dimensional simulations tailored to tackle specific problems. One example is that of 3D-RHD models of the solar atmosphere. But going deeper in the Sun, spatial and temporal scales become larger and longer, and the need for global models also becomes mandatory. The coupling between the convective envelope and the radiative interior, modeling the tachocline, the generation of magnetic fields, the propagation and angular momentum transfer by gravity waves are all problems that require this type of hydrodynamic and magnetohydrodynamic simulations. Of particular interest in this respect are \cite{rogers:2008,brun:2011,alvan:2014} and we encourage the reader to visit those references. 

\section{Final remarks}\label{sec:end}

Standard solar models have provided for many years a well defined reference for different fields of research, ranging from solar and stellar modeling to solar neutrinos and particle physics. The crisis caused by the reduction in the solar photospheric abundances of volatile elements has casted doubts on the soundness of SSMs as models that describe the global properties of the Sun accurately. While a definite solution to this problem still awaits, recent experimental results on radiative opacities of Fe allow us to be optimistic. But it is crucial that tests of solar structure that do not depend purely on opacities are carried out. This should be a fundamental aspect for solar neutrino experiments, with the goal of measuring CN neutrino fluxes with a precision not larger than 10\%. This would allow the determination of C+N solar core abundances to a precision comparable or better than those of spectroscopic technique in the solar photosphere. In addition to the solar composition problem, the possibility of comparing surface and core composition in the Sun would offer an unparalleled view on the efficiency of chemical mixing mechanisms in the solar interior and, by extension, in other low mass stars. 

\vskip .5cm

AS acknowledges support from the ESP2013-41268-R and ESP2014-56003-R (MINECO) and 2014SGR-1458 from the Generalitat de Catalunya.

%
%
%
%

\end{document}